\documentclass[preprint]{aastex631}

\begin{document}


\title{Can 1D radiative equilibrium models of faculae be used for calculating contamination of  transmission spectra?}


\correspondingauthor{Veronika Witzke}
\email{witzke@mps.mpg.de}

\author{Veronika Witzke}
\affiliation{Max Planck Institute for Solar System Research, Justus-von-Liebig-Weg 3,
37077 G\"ottingen, Germany}

\author{Alexander I. Shapiro}
\affiliation{Max Planck Institute for Solar System Research, Justus-von-Liebig-Weg 3,
37077 G\"ottingen, Germany}

\author{Nadiia~M.~Kostogryz}
\affiliation{Max Planck Institute for Solar System Research, Justus-von-Liebig-Weg 3,
37077 G\"ottingen, Germany}

\author{Robert Cameron}
\affiliation{Max Planck Institute for Solar System Research, Justus-von-Liebig-Weg 3,
37077 G\"ottingen, Germany} 

\author{Benjamin V.\ Rackham}
\altaffiliation{51 Pegasi b Fellow}
\affiliation{Department of Physics and Kavli Institute for Astrophysics and Space Research, Massachusetts Institute of Technology, Cambridge, MA 02139, USA}
\affiliation{Department of Earth, Atmospheric and Planetary Sciences, Massachusetts Institute of Technology, Cambridge, MA 02139, USA}

\author{Sara Seager}
\affiliation{Department of Physics and Kavli Institute for Astrophysics and Space Research, Massachusetts Institute of Technology, Cambridge, MA 02139, USA}
\affiliation{Department of Earth, Atmospheric and Planetary Sciences, Massachusetts Institute of Technology, Cambridge, MA 02139, USA}
\affiliation{Department of Aeronautics and Astronautics, MIT, 77 Massachusetts Avenue, Cambridge, MA 02139, USA}

\author{Sami K. Solanki}
\affiliation{Max Planck Institute for Solar System Research, Justus-von-Liebig-Weg 3,
37077 G\"ottingen, Germany}

\author{Yvonne~C.~Unruh}
\affiliation{Blackett Laboratory, Imperial College London, South Kensington Campus, London SW7 2AZ, UK. }

\begin{abstract}
%
The reliable characterization of planetary atmospheres with transmission spectroscopy requires realistic modeling of stellar magnetic features, since features that are attributable to an exoplanet atmosphere could instead stem from the host star's magnetic activity.
Current retrieval algorithms for analysing transmission spectra rely on intensity  contrasts of magnetic features from 1D radiative-convective models.
However, magnetic features, especially faculae, are not fully captured by such simplified models. 
Here we investigate how well such 1D models can reproduce 3D facular contrasts, taking a G2V star as an example. We employ the well established radiative magnetohydrodynamic code MURaM to obtain three-dimensional simulations of the magneto-convection and photosphere harboring a local small-scale-dynamo. Simulations without additional vertical magnetic fields are taken to describe the quiet solar regions, while simulations with initially 100\,G, 200\,G and 300\,G vertical magnetic fields are used to represent different magnetic activity levels. Subsequently, the spectra emergent from the MURaM cubes are calculated with the MPS-ATLAS radiative transfer code. 
We find that the wavelength dependence of facular contrast from 1D radiative-convective  models cannot reproduce facular contrasts obtained from 3D modeling. This has far reaching consequences for exoplanet characterization using transmission spectroscopy, where accurate knowledge of the host star is essential for
unbiased inferences of the planetary atmospheric properties.
\end{abstract}



\section{Introduction} 
\label{sec:intro}
In the past decade the transit method has become a highly productive tool for exoplanet hunting and characterization. 
During a transit, exoplanets appear larger at wavelengths where their atmosphere absorbs or scatters the radiation emitted by the host star. This allows identifying the composition of the planetary atmosphere by measuring a wavelength-dependent exoplanetary radius \citep[a so-called transmission spectrum,][]{seagerandsasselov2000, Brown2001, Charbonneauetal2002}. 
Transmission spectroscopy was given a huge boost by the advent of the James Webb Space Telescope (JWST), which has already been used to make the first definitive detection of carbon dioxide in an exoplanet atmosphere \citep{JWST_carbon_di} and even holds the promise  of studying atmospheres of rocky exoplanets \citep[e.g.,][]{Barstow_et_al2016, Morley_et_al2017, Lustig-Yaeger2019}. 

Apart from planetary atmospheres, signals in transmission spectra may result from surface inhomogeneities of host stars, caused by their magnetic activity. 
Namely,  magnetic features that remain unocculted during the transit induce a {\it wavelength-dependent} offset that affects the relative depth of transits and, consequently, the deduced planetary radii \citep[see, e.g.,][and references therein]{Rackham2018, Rackham2019}. 
For example, dark magnetic features outside of the transit chord lead the stellar disk to be darker on average than it is within the transit chord. This leads to increased relative transit depth and inferred planetary radii, which can be misinterpreted as absorption or scattering in the exoplanetary atmosphere. Similarly,  unocculted bright features decrease transit depths and thus can mask the genuine signal from the exoplanetary atmosphere. In many cases the signal from stellar  magnetic activity can significantly complicate the interpretation of transmission spectra \citep[see, e.g., the report from the NASA Exoplanet Exploration Program Analysis Group Study Analysis Group 21 for a detailed review;][]{SAG21}. Recent examples include \cite{Barclay_2021AJ}, who showed that the recently discovered water vapor signal on the Habitable-zone Sub-Neptune Exoplanet K2-18b \citep{Bennekeetal2019} can be equally well explained by the magnetic activity of the host star. Similarly, the analysis of WASP-103b's transmission spectrum by \cite{Kirketal2021} revealed strong evidence for magnetic activity of the host star ($4.3\sigma$)  and only weak evidence for absorption from the planetary atmosphere (with signals from H$_2$O, HCN, and TiO suggested at 1.9$\sigma$, 1.7$\sigma$, and 2.1$\sigma$, respectively).

The algorithms used to extract information from transmission spectra rely on the differences between the wavelength-dependence of the planetary signal and of the activity contamination in order to disentangle them. 
Until now, calculations of the wavelength-dependent activity contamination, which is an input of the retrieval algorithms, have represented spectra of magnetic features by radiative-equilibrium spectra of quiet stars with different effective temperatures \citep{Phoenix_grid}. Specifically, spots and faculae have been approximated by cooler and hotter stellar atmospheres, respectively \citep[see the detailed discussion in][]{SAG21}. One can expect that such approximations introduce significant inaccuracies. In particular, faculae are formed by small-scale magnetic concentrations  \citep{SolarB} and are heated by hot walls surrounding them \citep[see, e.g., Fig.~5 and its detailed discussion in the review by][]{ARA2013}. Thus, their brightness contrast is defined by 3D effects, which cannot be properly represented in one dimensional (1D) radiative equilibrium (RE) models.

Since accurate knowledge of the activity contamination and, in particular, its wavelength dependence  is essential for advancing exoplanet characterization, we examine the accuracy of 1D-approximated facular contrasts by comparing them to facular contrasts from three-dimensional (3D) radiative magnetohydrodynamic (MHD) simulations with the  MURaM code \citep{Vogler_Sch_2005, rempel_2014, rempel_2016}. Simulations of the solar atmosphere with the MURaM code have reached a high level of realism, reproducing highly detailed solar observations. They have been used to treat emergence of solar surface magnetic flux \citep{RemplelandCheung2014, Chen2017}, various magnetic features on the solar surface \citep{Vogler_Sch_2005, pores, spots}, as well as contribution of the convection \citep[see][]{Shapiro2017}, and small-scale magnetic concentrations \citep[see][]{Yeo2017} to solar brightness variability.

In this study we focus on a star with solar fundamental parameters and model faculae by initially adding a vertical magnetic field of 100\,G, 200\,G, and 300\,G, which corresponds to increasingly stronger levels of magnetic activity, to the state-of-the-art MURaM simulations of small-scale dynamo (SSD) by \cite{Tanayveer_2022} and \cite{Witzke_et_all_2022}. In Section~\ref{sec:model} we introduce the 3D approach for calculating facular contrasts and discuss the results. Subsequently, in Section~\ref{sec:results} we compare the 3D facular contrast to 1D models, focusing on whether it is possible to reproduce 3D results. In Section~\ref{sec:summary} we give a brief summary along with a discussion of future goals.


\section{3D modeling with the MURaM code} \label{sec:model}

The atmospheric structures for faculae and quiet regions are calculated using the `box-in-a-star' approach by employing the 3D radiative MHD code {MURaM}. %
MURaM solves the conservative  MHD equations for a compressible, partially ionized plasma to model the dynamics and energy transport in a Cartesian box. The radiative transfer (RT)  in the MURaM code is performed using a multi-group scheme with short characteristics \citep{nordlund_1982}. For the equation of state, pre-tabulated look-up tables generated by the FreeEOS code \citep{Irwin_freeeos_2012} are employed.  The simulated box covers 9 Mm $\times $ 9 Mm  (512 x 512 grid points) in the horizontal direction  and 5 Mm in depth with a resolution of 10 km. 


Our description of the quiet Sun is based on the most up-to-date 3D simulations as presented in \citet{Tanayveer_2022} and \citet{Witzke_et_all_2022}. One novel feature of these simulations is that they account for a small-scale dynamo operating near the stellar surface. The SSD leads to the formation of ubiquitous small-scale mixed polarity magnetic features  that  are always present at the solar surface, leading to considerable magnetic flux \citep[][]{Khomenko2003A&A,sveta2004, Trujillo2004Natur,Lites2004ApJ,Stenflo2013}. In our SSD simulations, the mean absolute value of the vertical magnetic field at the solar optical surface is 71 G  \citep{Witzke_et_all_2022}, which agrees well with solar observations \citep[see a detailed discussion in][]{rempel_2014, Rempel2020}. 
For the faculae we used the same setup as for the quiet regions including the SSD, but added initially vertical, unipolar and homogeneous magnetic fields of 100\,G, 200\,G, and 300\,G to achieve different levels of magnetic activity.
We note that in a previous study \citet{norris_phd} showed facular contrasts for different stellar types. However, at that time SSD simulations were not yet available and, thus, the quiet regions were modeled without any magnetic fields using hydrodynamic simulations.


%
For the spectra emergent from the 3D cubes,  we consider a spectral range of 200nm  to 2000nm.  
The calculations are performed at low spectral resolution using the MPS-ATLAS code \citep{mps-atlas-w2021}, which incorporates the new setup for opacity distribution functions by \cite{Miha2019} into a greatly updated version of the ATLAS9 code by \citet{Kurucz_2005_Atlas12_9}. The element abundance used for calculating the MURaM cubes as well as spectra emerging from the MURaM cubes are taken from \citet{Asplund_2009}. The  spectral synthesis is performed along rays with different viewing-angles, from $\mu =1.0$ (disk-center) to  $\mu =0.1$ (limb) in steps of 0.1 (where $\mu$ is the cosine of the angle between the observer’s line of sight and the local stellar radius).
 
For each of the four setups, that is the pure SSD setup representing the quiet Sun and the three setups with  added magnetic fields representing faculae, we store  a time series of  40 independent 3D cubes within a time interval of 10 solar hours. Subsequently, for each viewing angle, $\mu$, the spectra are first averaged over all rays in each cube and then over all cubes in the time series. Then we define  the relative facular contrast as 
\begin{equation}
\label{eq:contrast3d}
    C_B (\lambda, \mu)  = (I_B (\lambda, \mu) - I_{QS}(\lambda, \mu) ) / I_{QS}(\lambda, \mu),
\end{equation}
where $I_{QS}( \lambda, \mu)$ is spectral intensity from the quiet Sun (represented by the SSD setup) and $I_B (\lambda, \mu)$ (with $B=100 \,{\rm G}, 200 \, {\rm G}, 300 \,{\rm G} $) are from setups representing faculae.

Blue and black lines in Figure~\ref{fig:01fig} show the spectral dependence of the relative facular contrasts at the disk-center ($\mu = 1.0$) and  limb ($\mu = 0.1$) as well as disk-integrated values.
One can see that the contrasts strongly increase from the disk-center to the limb.
The contrasts change by several orders of magnitude from the ultraviolet (UV) to the infrared (IR),  with disk-center contrasts even becoming negative at some wavelengths (blue parts of curves shown in Figure~\ref{fig:01fig}) so that we plot the absolute values. 

With increasing magnetic activity, the facular regions become dark at disk-center, first in the infrared (IR)  then also in the visible  (see middle panel in Fig.~\ref{fig:01fig}).
Strikingly, the disk-integrated and disk-center contrasts show a very complex and non-monotonous wavelength dependence. At the limb, however, the contrast does not show strongly pronounced line features and appears smoother. This is because of the effect of spectral lines \citep[for a detailed discussion see][]{Shapiro2015}, that become  progressively more important towards the disk-center.



\section{Comparison of contrasts from 1D and 3D models}  \label{sec:results}
To compare the facular contrasts presented in Sect.~\ref{sec:model} to contrasts obtained by 1D modeling, we generate 1D atmospheric structures under the assumption of radiative-convective equilibrium with the treatment of the convection described by mixing length theory \citep{ML}. 
Subsequently we calculate the specific intensities for the same viewing-angles as used in the 3D approach. We note that the 1D calculations using the MPS-ATLAS code have been extensively validated against available observations and models of quiet solar/stellar atmospheres \citep[see][]{mps-atlas-w2021, Nadiia2022}. Moreover, we compare the 3D contrasts to the state-of-the-art practice in exoplanetary atmospheric retrievals, namely, contrasts derived from the PHOENIX spectral grid \citep{Phoenix_grid}.

In our first experiment effective temperatures, $\rm T_{eff}$, of the 1D atmospheric structures are chosen so as to produce the same disc integrated bolometric intensity outputs as those of the 3D simulations averaged over a period of circa 10 hours of solar time.
These are $\rm 5787 \, K$, $\rm 5805 \, K$, $\rm 5823 \, K$, and $\rm 5835 \, K$ for 
the SSD, 100G, 200G, and 300G setups, respectively.  Consequently, 1D atmospheres with the above effective temperatures are considered to represent faculae of different strength or intensity. The other fundamental parameters such as the surface gravity  ($\log g = 4.3 ) $ and metallicity ($\rm M/H = 0.0$) are  kept the same as in the MURaM simulations. For the contrasts derived from the PHOENIX grid\footnote{http://phoenix.astro.physik.uni-goettingen.de}, we interpolated the disk-integrated flux and the specific intensities for disk-center to the desired effective temperature and surface gravity from the closest available stellar parameters. We did not consider calculations for the limb, because specific intensities in the PHOENIX grid are calculated for different viewing-angle grids depending on stellar parameters, which makes interpolating even more inaccurate.
%
The facular contrasts are then calculated employing Eq.~\ref{eq:contrast3d},  using spectra emergent from the  1D models with the effective temperatures given above. 

\begin{figure}[ht!]
\plotone{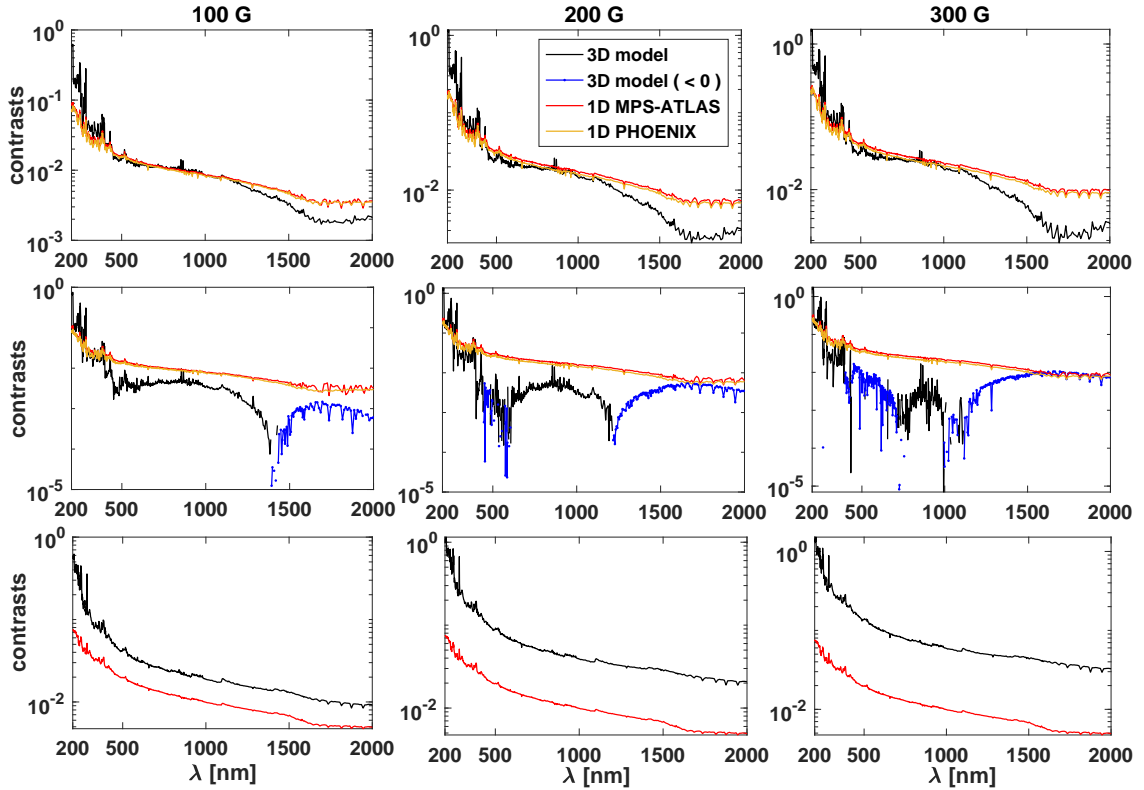}
\caption{Relative facular contrast from 3D simulations (as given in Eq.~\ref{eq:contrast3d}) and from 1D models (as discussed in Section~\ref{sec:results}) for different magnetic activity levels (left to right columns) and viewing angles (top to bottom rows). The top row shows the relative contrast for the disk-integrated flux, middle row for  the disk-center, and bottom row for the limb ($\mu = 0.1$).  
\label{fig:01fig}}
\end{figure}

The 1D contrasts calculated this way are shown in Figure~\ref{fig:01fig} (red and orange lines).  It becomes evident that at  disk-center and integrated over the disc, the wavelength dependence of the 3D and 1D contrasts are substantially different: 
In particular, 1D modeling greatly overestimates the contrast in the visible and IR, but underestimates it in the UV compared to contrast modeled in 3D (see top and middle panels of Figure~\ref{fig:01fig}). This implies that the disk-center and disk-integrated facular contrasts cannot be even remotely approximated by 1D radiative equilibrium  models. There is only few and small difference between the MPS-ATLAS contrasts and the PHOENIX contrast. Thus, at disk-center PHOENIX contrasts in the IR display fewer line features compared to MPS-ATLAS contrasts.

At the same time, the wavelength dependences of the 1D and 3D contrasts at the limb are similar (see bottom panels of Figure~\ref{fig:01fig}), but 1D facular contrasts  are substantially lower than contrasts given by the 3D simulations.  
Consequently,  facular contrasts at the limb can be approximated reasonably by 1D modeling  but with significantly overestimated facular temperatures (since an increase of the facular temperature will lead to a scaling of the facular contrast with a factor only slightly dependent on wavelength, see below). 

This result can be understood by recalling that faculae are caused by  ensembles of small-scale magnetic concentrations. These concentrations can be well represented by flux tubes \citep[see reviews by][]{SolarB,ARA2013}. The temperature structure within these flux tubes is strongly affected by the radiative heating from the hot walls and, thus, cannot be approximated by a radiative-convective equilibrium 1D model. Consequently, these models dramatically fail to describe the wavelength dependence of facular contrast at the disk center and intermediate disk positions. The emergent intensity of faculae at the limb comes from the hot walls themselves, which can be reasonably approximated by 1D radiative-convective models.




The retrieval algorithms used for the analysis of transmission spectra usually keep the surface coverage fraction of stellar magnetic features as a free parameter which is constrained during the fitting of the transmission spectra \citep[see, e.g.,~a detailed discussion in][]{Kirketal2021}. Therefore only the {\it wavelength dependence} of the facular contrast is important for disentangling between planetary and facular signatures in transmission spectra --- any relative offset in facular contrast can be compensated by scaling the surface coverage fraction of faculae without affecting the retrieved planetary signal. 

In this context, we conduct our second experiment to investigate whether 1D models with arbitrary chosen temperatures (in contrast to the first experiment shown in Figure~\ref{fig:01fig}, where we took temperatures from the 3D simulations) can reproduce the wavelength dependence of facular contrast in spectral intervals that are often used in transmission spectroscopy \citep[see, for example,][]{Wakeford2020, Lothringer2022Natur}, namely  $\rm 200 - 400 \, nm$, $\rm 200 - 800 \, nm$, and $\rm 1100- 1700 \, nm$. For this we represent the quiet Sun by the 1D model with effective temperature $\rm T_{\rm QS} = 5790 K$ and faculae by 1D models with temperatures $\rm T_{\rm QS} + \Delta T$, where we consider  $\rm \Delta T = 20 \, {\rm K}, 50 \, {\rm K}, 100 \, {\rm K}, 200 \, {\rm K}, 300 \, {\rm K}$. We note that the exact choice of 
$\rm T_{QS}$ plays little role in our experiment since the intensity contrasts mainly depend on $ \rm \Delta T$.

\begin{figure}[ht!]
\plotone{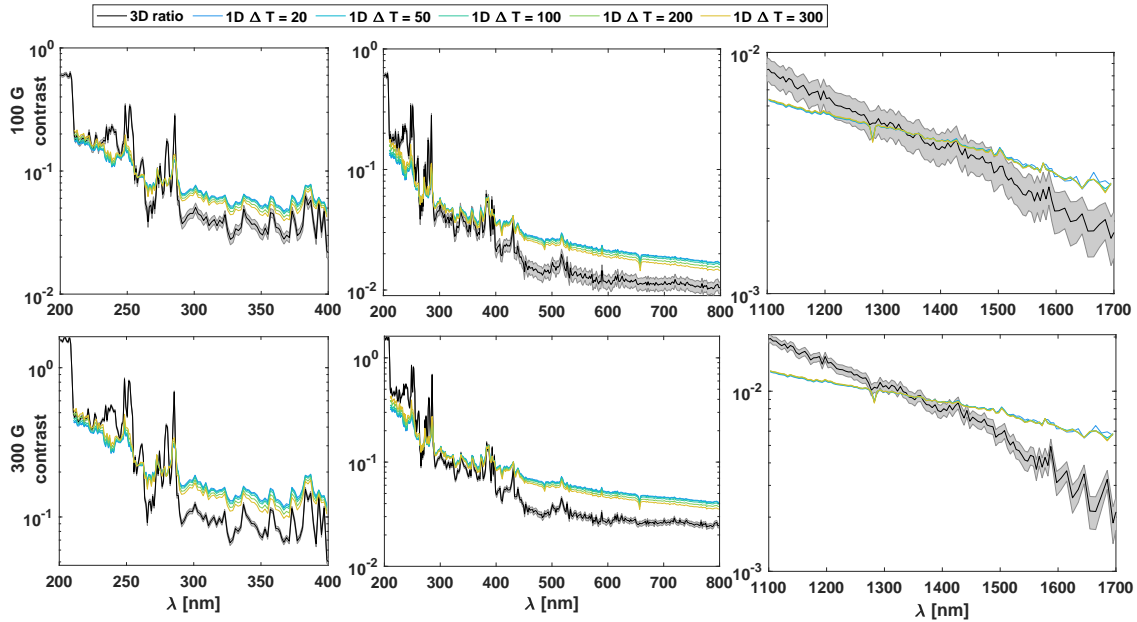}
\caption{Disk-integrated relative facular contrast from 3D simulations (as given in Eq.~\ref{eq:contrast3d}) and normalized (see text for details) relative contrasts from 1D models for different $\rm \Delta T$. The gray shaded area corresponds 
to the standard deviation of the scatter in contrasts due to 
temporal fluctuations in the 3D simulations.  Different levels of magnetic activity (100\,G in the top row and 300\,G in the bottom row)  and different wavelength intervals (left to right columns) are shown.
\label{fig:02fig}} 
\end{figure}

In each of the three wavelength intervals, we normalize the 1D contrast, so that the integrals of 1D and 3D contrasts over the corresponding wavelength interval are equal. The result is presented in Figure~\ref{fig:02fig}, which shows the normalized 1D disk-integrated contrasts for different $\rm \Delta T$ values and corresponding 3D contrasts for different magnetizations. The contrast below 210\,nm has a steep increase due to the Al I ionization edge. Thus, we excluded the region 200\,nm to 210\,nm from normalization and show 1D contrasts only longwards of 210\,nm.  
%
%

Figure~\ref{fig:02fig} shows that the wavelength dependence of the 1D contrasts is only barely affected by the value of  $\rm \Delta T$. In particular, they share all spectral features in the UV and in the visible.  1D RE models result in a totally different  overall slope compared to the 3D contrast even in the relatively narrow spectral intervals shown in  Figure~\ref{fig:02fig}. We note that, the overall slope is mainly determined by the behavior of the contrast in continuum.
%
%
%
Furthermore, one can see that 1D calculations dramatically fail to capture the sophisticated wavelength dependence of facular contrast caused by spectral lines and revealed by the 3D calculations. The more pronounced lines in the 3D contrasts can be attributed to different vertical temperature gradients in the photosphere and to the temperature inhomogeneities due to granulation and magnetic fields.

\section{Implications and future goals } \label{sec:summary}
We showed that facular contrast calculated using 1D models fail to reproduce the complex wavelength dependence that is found using 3D models. Furthermore, contrasts from 1D radiaitve equilibrium models cannot be fudged by adjusting the fraction of faculae present on the surface to get a closer match with contrasts from 3D models. 

This result has consequences for transmission spectroscopy of transiting exoplanets, which is a widely used technique to deduce information on the structure and composition of exoplanet atmospheres.  Conclusions drawn from transmission spectroscopy of transiting exoplanets, in particular, those relying on UV and blue optical data of planets transiting active stars, are potentially inaccurate, as the transmission spectra may be compromised by inhomogeneities on the host star, which so far were modeled with 1D radiaitve equilibrium models \citep{Espinoza_2019, Kirketal2021, Welbanks_2022}.
%

In this work we focused on the solar case to analyse if contrasts from 1D radiaitve equilibrium  models can reproduce contrasts from 3D MHD simulations. Since we find that it is essential to use 3D models for active region modeling, in forthcoming publications we will extend our calculations of 3D facular contrasts from the solar fundamental parameters to other main-sequence stars of different metallicity and spectral type using the simulations of \cite{Tanayveer_2022} and \cite{Witzke_et_all_2022}.  
We plan to complement this work by computing spot contrasts using simulations by \citet{Mayukh2020}.


\section{Acknowledgments}
\begin{acknowledgments}
This work has received funding from the European Research Council (ERC) under the European Union's Horizon 2020 research and innovation programme (grant agreement No. 715947).  This work has been partially supported by the Max Planck Society's grant ``PLATO Science" and DLR's PLATO grant Nr.~$50$OO$1501$ and $50$OP$1902$. B.V.R.~thanks the Heising-Simons Foundation for support.  
\end{acknowledgments}

\bibliography{biblist}{}
\bibliographystyle{aasjournal}



\end{document}